\documentstyle[psfig,referee]{laa}
\thesaurus{12.03.4; 12.12.1; 11.06.1}
\title{Tidal torques and the clusters of galaxies evolution} 
\author{A. Del Popolo\inst{1,2}, and ~ M. Gambera\inst{1,3}}  
\offprints{M. Gambera E-mail: mga@sunct.ct.astro.it}
\institute{$^1$ Istituto di Astronomia dell'Universit\`a di Catania, 
Viale A.Doria, 6 - I 95125 Catania, ITALY \\
$^2$ Facolt\`{a} di Ingegeneria, Universit\`{a} Statale di Bergamo, Piazza Rosate, 2 - I 24129 Bergamo, ITALY\\
$^3$ Osservatorio Astrofisico di Catania and CNR-GNA, 
Viale A.Doria, 6 - I 95125 Catania, ITALY 
}
\date{}
\begin{document}
\maketitle
\begin{abstract}
We study the effect of tidal torques on the collapse of density peaks through the equations of motion of a shell of barionic matter falling
into the central regions of a cluster of galaxies. We calculate the time of
collapse of the perturbation taking into account the gravitational
interaction of the quadrupole moment of the system with the tidal field of
the matter of the neighbouring proto-clusters. We show that within
high-density environments, such as rich clusters of galaxies, tidal torques
slow down the collapse of low-$\nu $ peaks producing an observable
variation in the time of collapse of the shell and, as a consequence, a
reduction in the mass bound to the collapsed perturbation. Moreover, the delay
of the collapse produces a tendency for less dense regions to accrete less
mass, with respect to a classical spherical model, inducing a {\it biasing} of
over-dense regions toward higher mass. Finally we calculate the bias
coefficient using a selection function properly defined showing that for a
Standard Cold Dark Matter (SCDM) model this ${\it bias}$ can account for a 
substantial part of the total bias required by observations on cluster scales. 
\end{abstract}

\section{Introduction}

It has long been
speculated on the fundamental role that the angular momentum could play 
in determining the fate of collapsing proto-structures and several models 
have been proposed to correlate the galaxy type with the angular momentum 
per unit mass of the structure itself (Faber 1982; Kashlinsky 1982; Fall 1983).
 Some authors (see
Barrow \& Silk 1981; Szalay \& Silk 1983 and Peebles 1990) have proposed that
non-radial motions would be expected within a developing proto-cluster due to
the tidal interaction of the irregular mass distribution around
them, typical of hierarchical clustering models, with the
neighbouring proto-clusters.
The kinetic energy of these non-radial motions prevents the collapse
of the proto-cluster, enabling the same to reach statistical equilibrium
before the final collapse (the so-called previrialization conjecture by
Davis \& Peebles 1977, Peebles 1990). This effect may prevent the
increase in the slope of the mass autocorrelation function at
separations given by $\xi(r,t) \simeq 1$, expected in the scaling solution for
the rise in $\xi(r,t)$ but not observed in the galaxy two-point
correlation function. The role of non-radial motions has been 
pointed by several authors (see Davis \& Peebles 1983: Gorski 1988; 
Groth et al. 1989; Mo et al. 1993; van de Weygaert \& Babul 1994; Marzke 
et al. 1995 and Antonuccio-Delogu \& Colafrancesco 1995). 
Antonuccio-Delogu \& Colafrancesco 
derived the conditional probability distribution $f_{pk}({\bf v}| \nu)$  
of the peculiar velocity around a peak of a Gaussian density field and 
used the moments of the velocity distribution to study the 
velocity dispersion around the peak.
They showed that regions of the proto-clusters at radii, $ r$, 
greater than the filtering length, $ R_{f}$, contain predominantly 
non-radial motions. \\
Non-radial motions change the energetics of the collapse model
by introducing another
potential energy term. In other words one expects that non-radial motions
change the characteristics of the collapse and in particular the
{\it turn around} epoch, $t_{m}$,
and consequently the critical threshold, $ \delta_{c}$, for collapse.
Here, we want to remind that $ t_{m}$ is the time at which the linear density
fluctuations, that generate the cosmic structures, detach
from the Hubble flow. The turn-around epoch is given by: 
\begin{equation}
t_{m} = \left[\frac{ 3 \pi}{32 G \rho_{b}} ( 1 +\overline{\delta})
\right]^{1/2} (1+z)^{3/2}
\end{equation}
where $ \rho_{b} $ is the mean background density, $z$ is the redshift and
$\overline{\delta}$ is the mean over-density within the non-linear region. 
After the {\it turn around} epoch, the fluctuations start to recollapse.
As known for a spherical top hat model, the perturbation of the density
field is completely collapsed when
\begin{equation}
\overline{\delta} = \delta_{c}=(3/5)(\frac{3 \pi T_{c}}{4 t_{m}})^{2/3} = 1.68
\end{equation}
where $T_{c}$ is the time of collapse which is twice the turn around epoch.
One expects that non-radial motions produce firstly
a change in the turn around epoch, secondly a new functional
form for $ \delta_{c}$, thirdly
a change in the mass function calculable with the Press-Schechter
(1974) formula and finally a modification of the two-point correlation function. As we shall show in a forthcoming paper (Del Popolo \& Gambera 1997b) 
non-radial motions can reduce several discrepancies between the 
SCDM model and observations: the strong
clustering of rich clusters of galaxies ($\xi _{cc}(r) \simeq (r/25h^{-1}Mpc) ^{-2}$)  far in excess of CDM predictions (Bahcall \&
Soneira 1983), the X-ray temperature distribution function of clusters 
over-producing the observed cluster abundances (Bartlett \& Silk 1993).\\
For the sake of completeness, we remember that alternative models with more large-scale power than SCDM have been introduced
in order to solve the latter problem. Several authors (Peebles 1984;
Efstathiou et al. 1990; Turner 1991; 
White et al. 1993) have lowered the matter density under
the critical value ($%
\Omega _m<1$) and have added a cosmological constant in order to retain
a flat universe ($\Omega _m+\Omega _\Lambda =1$) .
The spectrum of the matter density is specified by the
transfer function, but its shape is
affected because of the fact that the epoch of
matter-radiation equality is earlier,
$1+z_{eq}$ being increased by a factor $1/\Omega_{m}$. 
Around the epoch $z_\Lambda $ the growth of the
density contrast slows down and ceases after $z_\Lambda $.
As a consequence the
normalisation of the transfer function begins to fall, even if its shape
is retained.\\ 
Mixed dark matter models (MDM) (Bond et al. 1980; Shafi \& Stecker
1984; Valdarnini \& Bonometto 1985; Holtzman 1989;
Schaefer 1991; Schaefer \& Shafi 1993; Holtzman \& Primack 1993) increase
the large-scale power because free-streaming neutrinos damp the power on
small scales. Alternatively changing the primeval spectrum
several problems of SCDM are
solved (Cen et al. 1992). Finally, it is possible to assume
that the threshold for galaxy formation
is not spatially invariant but weakly modulated ($2\%-3\%$ on scales $%
r>10h^{-1}Mpc$) by large scale density fluctuations, with 
the result that the clustering on
large-scale is significantly increased (Bower et al. 1993). \\
Moreover, this study of the role of non-radial motions in
the collapse of density perturbations can help us to give a deeper insight
in to the so-called problem of biasing. As pointed out by Davis et al. (1985),
unbiased CDM presents several problems: pairwise velocity dispersion
larger than the observed one, galaxy correlation function steeper than
that observed (see Liddle \& Lyth 1993 and Strauss \& Willick 1995). 
The remedy to these problems is the concept of biasing (Kaiser 1984), 
i.e. that galaxies are more strongly clustered than
the mass distribution from which they originated.
The physical origin of such biasing is not yet clear even if several
mechanisms have been proposed (Rees 1985; Dekel \& Silk 1986; Dekel \& 
Rees 1987; Carlberg 1991; Cen \& Ostriker 1992; Bower et al. 1993; Silk \& 
Wyse 1993). Recently Colafrancesco, Antonuccio-Delogu \& Del Popolo (1995, 
hereafter CAD) and Del Popolo \& Gambera (1997a) 
have shown
that dynamical friction delays the collapse of low-$\nu$ peaks inducing a 
bias of dynamical nature. Because of the dynamical friction, under-dense 
regions in clusters (the clusters outskirts) accrete less mass than  
that accreted in absence of this dissipative effect and as a consequence 
over-dense regions are biased toward higher mass (Antonuccio-Delogu \& Colafrancesco 
1994 and Del Popolo \& Gambera, 1996).
Non-radial motions acts in a 
similar way to dynamical friction: they delay the shell collapse
consequently inducing a dynamical bias similar to that produced by dynamical
friction. This dynamical bias can be evaluated defining a selection function
similar to that given in CAD and
using Bardeen, Bond, Szalay \& Kaiser (1986, hereafter BBKS) prescriptions. \\
 The methods used in this paper are fundamentally some results
of the statistics of Gaussian random fields, the biased galaxy
formation theory and the spherical model for the collapse of density perturbations.
In particular, we calculate the specific angular momentum acquired by protoclusters and the time of
collapse of protoclusters
using the Gaussian random fields theory and the spherical collapse
model following Ryden's (1988a, hereafter R88a) approach.
The selection function that we introduce is general and obtained by the
only hypothesis of Gaussian density field. The approach and the
final result is totally different from BBKS selection function and similar to
that of Colafrancesco, Antonuccio \& Del Popolo (1995).
Only the biasing parameter is obtained from a BBKS approximated formula.
This choice will be clarified in the following sections of the paper.\\ 
The plan of the paper is the following: in Sect.  2 we obtain the total
specific angular momentum acquired during expansion by a proto-cluster.
In Sect.  3 we use the calculated specific angular momentum to obtain the
time of collapse of shells of matter around peaks of density having
$\nu_c = 2, 3, 4$ and we compare the results with Gunn \& Gott's 
(1972, hereafter GG) spherical
collapse model. In Sect.  4 we derive a selection function for the peaks
giving rise to proto-structures while in Sect.  5 we calculate some values for 
the bias parameter, using the selection function derived, on three relevant 
filtering scales. 
Finally in Sect.  6 we discuss the results obtained.

\section{Tidal torques}

The explanation of galaxies spins gain through tidal torques was pioneered by
Hoyle (1949) in the context of a collapsing protogalaxy. Peebles (1969)
considered the process in the context of an expanding world model showing
that the angular momentum gained by the matter in a random comoving {\it %
Eulerian} sphere grows at the second order in proportion to $t^{5/3}$ (in a
Einstein-de Sitter universe), since the proto-galaxy was still a small
perturbation, while in the non-linear stage the growth rate of an oblate
homogeneous spheroid decreases with time as $t^{-1}$.\\
More recent analytic computations (White 1984; Hoffman 1986 and R88a) 
and numerical simulations (Barnes \& Efstathiou 1987) 
have re-investigated the role of tidal torques in originating 
galaxies angular momentum. In particular White (1984) 
considered an analysis by Doroshkevich (1970) that showing as 
the angular momentum of galaxies grows to first order in proportion to $ t$ 
and that Peebles's result is a consequence of the spherical symmetry imposed to 
the model. White showed that the angular momentum of a Lagrangian sphere does
not grow either in the first or in the second order while the angular 
momentum of a non-spherical volume grows to the first order in 
agreement with Doroshkevich's result. \\
Hoffman (1986) has been much more involved in the analysis of the
correlation of the growth of angular momentum with the density perturbation $
\delta(r) $. He found an angular momentum-density anticorrelation: high density
peaks acquire less angular momentum than low density peaks. One way to study
the variation of angular momentum with radius in a galaxy is that followed
by R88a. In this approach the protogalaxy is divided into a series
of mass shells and the torque on each mass shell is computed separately. The
density profile of each proto-structure is approximated by the superposition
of a spherical profile, $\delta (r)$, and a random CDM distribution, ${\bf %
\varepsilon (r)}$, which provides the quadrupole moment of the protogalaxy.
To the first order, the initial density can be represented by:
\begin{equation}
\rho ({\bf r})=\rho _b\left[ 1+\delta (r)\right] \left[ 1+\varepsilon ({\bf 
r})\right] 
\end{equation}
where $\rho_{b}$ is the background density and $ \varepsilon(\bf r)$
is given by:
\begin{equation}
\langle |\varepsilon _k|^2 \rangle = P(k)
\end{equation}
being $ P(k)$ the power spectrum, while the density profile 
is (Ryden \& Gunn 1987):
\begin{equation}
\langle \delta (r) \rangle =\frac{\nu \xi (r)}{\xi (0)^{1/2}}-\frac{\vartheta (\nu
\gamma ,\gamma )}{\gamma (1-\gamma ^2)}\left[ \gamma ^2\xi (r)+\frac{%
R_{\ast }^2}3\nabla ^2\xi \right] \cdot \xi (0)^{-1/2} 
\label{eq:dens}
\end{equation}
where $\nu $ is the height of a density peak, $\xi (r)$ is the two-point
correlation function, $\gamma $ and $R_{\ast}$  are two spectral parameters 
(BBKS, Eq. 4.6a, 4.6d) while $ \vartheta (\gamma \nu ,\gamma )$ is a function 
given in BBKS (Eq. 6.14).
As shown by R88a the net rms torque on a mass shell centered on the
origin of internal radius $ r$ and thickness $\delta r$ is given by:
\begin{equation}
\langle|\tau |^2\rangle^{1/2}=\sqrt{30}\left( \frac{4\pi }5G\right) \left[
\langle a_{2m}(r)^2 \rangle \langle q_{2m}(r)^2 \rangle - \langle 
a_{2m}(r)q_{2m}^{*}(r)\rangle^2\right] ^{1/2}
\label{eq:tau}
\end{equation}
where $q_{lm}$, the multipole moments of the shell and $a_{lm}$, the tidal
moments, are given by:
\begin{equation}
\langle q_{2m}(r)^2 \rangle =\frac{r^4}{\left( 2\pi \right) ^3}M_{sh}^2 
\int k^2dkP\left(k\right) j_2\left( kr\right) ^2
\end{equation}
\begin{equation}
\langle a_{2m}(r)^2 \rangle = \frac{2\rho _b^2r^{-2}}\pi \int dkP
\left( k\right) j_1\left(kr\right) ^2
\end{equation}
\begin{equation}
\langle a_{2m}(r)q_{2m}^{*}(r) \rangle =\frac r{2\pi ^2}\rho _bM_{sh}
\int kdkP\left(k\right) j_1\left( kr\right) j_2(kr)
\end{equation}
where $ M_{sh}$ is the mass of the shell, $j_1(r) 
$ and $j_2(r)$ are the spherical Bessel function of first and second order
while the power spectrum $ P(k)$ is given by:
\begin{eqnarray}
P(k)&=&Ak^{-1}[\ln \left( 1+4.164k\right)]^2 \cdot (192.9+1340k+ \nonumber \\
& + &  1.599\cdot 10^5k^2+1.78\cdot 10^5k^3+3.995\cdot
10^6k^4)^{-1/2}
\end{eqnarray}
(Ryden \& Gunn 1987). The normalization constant $ A$ can be obtained, 
as usual, by fixing that the mass variance at $8h^{-1}Mpc$, that is 
$\sigma _{8}$, be equal to unity.
Filtering the spectrum on cluster scales, $R_{f}=3h^{-1} Mpc$, 
we have obtained the rms torque, $%
\tau (r)$, on a mass shell using Eq. (\ref{eq:tau}) then we 
obtained the total specific
angular momentum, $h(r,\nu)$, acquired during expansion integrating the torque over
time (R88a Eq. 35):
\begin{equation}
h(r,\nu)=  \frac{\tau_o t_o \overline{\delta}_o^{-5/2}}{\sqrt[3]{48}M_{sh}}%
\int_0^\pi \frac{\left( 1-\cos \theta \right) ^3}{%
\left( \vartheta -\sin \vartheta \right) ^{4/3}}\frac{f_2(\vartheta) \cdot%
d\vartheta}{f_1(\vartheta )-f_2(\vartheta )\frac{\delta _o}%
{\overline{\delta _o}}}
\label{eq:ang}
\end{equation}
where the functions $f_1(\vartheta )$, $f_2(\vartheta )$ are given by R88a 
(Eq. 31) while the mean over-density inside the shell, 
$\overline{\delta} (r)$,
is given by (R88a): 
\begin{equation}
\overline{\delta}(r,\nu)=\frac{3}{r^{3}} \int_{0}^{\infty} d\sigma 
\sigma^{2} \delta(\sigma)
\end{equation}
In Fig. 1 we show the variation of 
$h(r,\nu)$ with the distance $r$ for three values of the peak height 
$\nu $. The
rms specific angular momentum, $h(r,\nu)$, increases with distance $r$ while peaks
of greater $\nu $ acquire less angular momentum via tidal torques.
\begin{figure}[ht]
\psfig{file=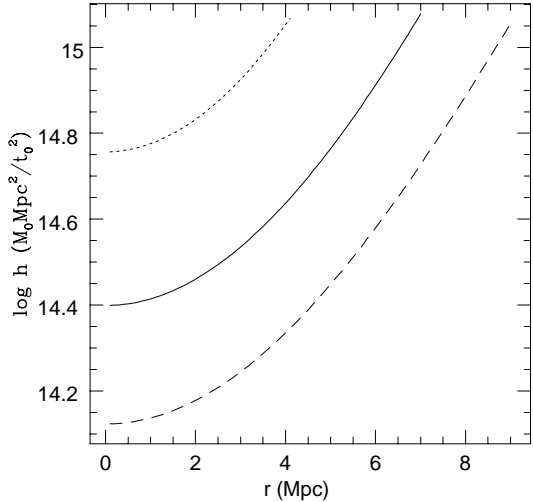,width=12cm}
\caption[]{The specific angular momentum, in units of $M_{\odot}$, Mpc and 
the Hubble time, $t_{0}$, for
three values of the parameter $\nu$ ($\nu=2$ dotted line, $\nu=3$
solid line, $\nu=4$ dashed line)  and for $R_{f}=3h^{-1}Mpc$.}
\end{figure}
This is the angular momentum-density anticorrelation showed by Hoffman
(1986). This effect arises because the angular momentum is proportional to
the gain at turn around time, $t_{m}$, which in turn is proportional
to $\overline{\delta}(r,\nu)^{-\frac 32} \propto \nu^{-3/2}$.

\section{Shell collapse time}

One of the consequences of the angular momentum acquisition by a mass shell
of a proto-cluster is the delay of the collapse of the proto-structure. As
shown by Barrow \& Silk (1981) and Szalay \& Silk (1983) the gravitational
interaction of the irregular mass distribution of proto-cluster with the
neighbouring proto-structures gives rise to non-radial motions, within the
protocluster, which are
expected to slow the rate of growth of the density contrast and to delay or
suppress the collapse. According to Davis \& Peebles (1977) the kinetic energy
of the resulting non-radial motions at the epoch of maximum expansion
increases so much to oppose the recollapse of the proto-structure. Numerical
N-body simulations by Villumsen \& Davis (1986) showed a tendency to
reproduce this so-called previrialization effect. In a more recent paper by
Peebles (1990) the slowdown in the growth of density fluctuations and the
collapse suppression after the epoch of the maximum expansion were re-obtained
using a numerical action method.\\
In the central regions of a density peak ($r\leq 0.5R_f$) the 
velocity dispersion attains nearly the same value (Antonuccio-Delogu \& 
Colafrancesco 1995) while at larger radii ($r \geq R_f$) the radial 
component is lower
than the tangential component. This means that motions in the outer regions
are predominantly non-radial and in these regions the fate of the infalling
material could be influenced by the amount of tangential velocity relative
to the radial one. This can be shown writing the equation of motion of a
spherically symmetric mass distribution with density $n(r)$ (Peebles 1993):
\begin{equation}
\frac \partial {\partial t}n \langle v_r \rangle +\frac \partial {\partial
r}n \langle v_r^2 \rangle +\left( 2 \langle v_r^2 \rangle - \langle 
v_\vartheta ^2 \rangle \right) \frac nr+n(r)\frac \partial {\partial t}
\langle v_r \rangle = 0 
\label{eq:peeb}
\end{equation}
where $ \langle v_r \rangle $ and $ \langle v_\vartheta \rangle $ 
are, respectively, the mean radial and tangential streaming velocity. 
Eq. (\ref{eq:peeb}) shows that high tangential velocity dispersion 
$(\langle v_\vartheta ^2 \rangle \geq  2 \langle v_r^2 \rangle)$ 
may alter the infall pattern. The expected delay in the collapse of 
a perturbation may be calculated solving the equation for the 
radial acceleration (Peebles 1993):
\begin{equation}
\frac{dv_r}{dt}=\frac{L^2(r,\nu)}{M^{2} r^3}-g(r) \label{eq:coll}
\end{equation}
where $L(r,\nu)$ is the angular momentum and $g(r)$ the acceleration.
Writing the proper radius of a shell in terms of the expansion parameter, 
$a(r_i,t)$:
\begin{equation}
r(r_i,t)=r_i a(r_i,t)
\end{equation}
remembering that
\begin{equation} 
M=\frac{4\pi }3\rho_{b} (r_i,t)a^3(r_i,t)r_i^3
\end{equation}
and that $ \rho_{b} = \frac{3H_{0}^{2}}{8 \pi G}$, where $ H_0$ is the 
Hubble constant and assuming that no shell crossing occurs so 
that the total mass inside each shell remains constant, that is:
\begin{equation} 
\rho (r_i,t)=\frac{\rho _i(r_i,t)}{a^3(r_i,t)} 
\end{equation}
Eq. (\ref{eq:coll}) may be written as:
\begin{equation}
\frac{d^2a}{dt^2}=-\frac{H^2(1+\overline{\delta} )}{2a^2}+\frac{4G^2L^2}{H^4(1+
\overline{\delta}
)^{2} r_i^{10}a^3} 
\label{eq:sec}
\end{equation}
or integrating the equation once more:
\begin{equation}
\left(\frac{da}{dt}\right)^2 = H_i^2\left[ \frac{1+\overline{\delta} }{a} 
\right] +\int 
\frac{8G^2L^2}{H_i^2r_i^{10}\left( 
1+\overline{\delta} \right) ^2} \frac 1{a^3} da-2C 
\label{eq:ses}
\end{equation}
where $C$ is the binding energy of the shell. The value of 
$ C$ can be obtained using the condition for turn around $ \frac{da}{dt} = 0$ 
when $ a = a_{max}$ leading to the new equation:
\begin{equation}
\left(\frac{da}{dt}\right)^2 = H_i^2\left[ \frac{1+\overline{\delta} }{a}-
\frac{1+\overline{\delta}}{a_{max}} 
\right] +\int_{a_{max}}^{a} 
\frac{8G^2L^2}{H_i^2r_i^{10}\left( 
1+\overline{\delta} \right)^2} \frac {1}{a^{`3}}d a^{`}
\end{equation}
Equation (\ref{eq:coll}) or equivalently Eq. (\ref{eq:sec}) 
may be solved using the initial conditions: $ (\frac{da}{dt}) = 0$,
$ a = a_{max} \simeq 1/ \overline{\delta}$ and
using the function $h(r,\nu) = L(r,\nu)/M_{sh}$
found in Sect. 2 to obtain the time
of collapse, $T_c(r, \nu)$. \\ 
In Figs. 2  $\div$  4 we compare the results for 
the time of collapse, $T_c$, for $\nu =2,3,4$ with the 
time of collapse of the classical GG spherical model:
\begin{equation}
T_{c0}(r,\nu)=\frac \pi {H_i} [\overline{\delta}(r,\nu)]^{-3/2}
\end{equation}
\begin{figure}[ht]
\psfig{file=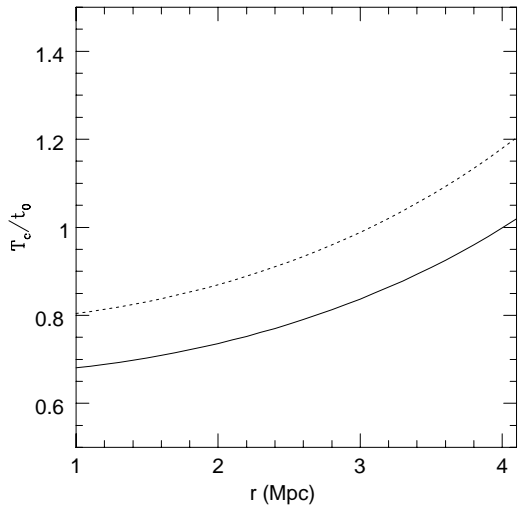,width=12cm}
\caption[]{The time of collapse of a shell of matter in units of the age of the
Universe $t_{0}$ for $\nu=2$ (dotted line) compared with GG's
model (solid line).}
\end{figure}
\begin{figure}[ht]
\psfig{file=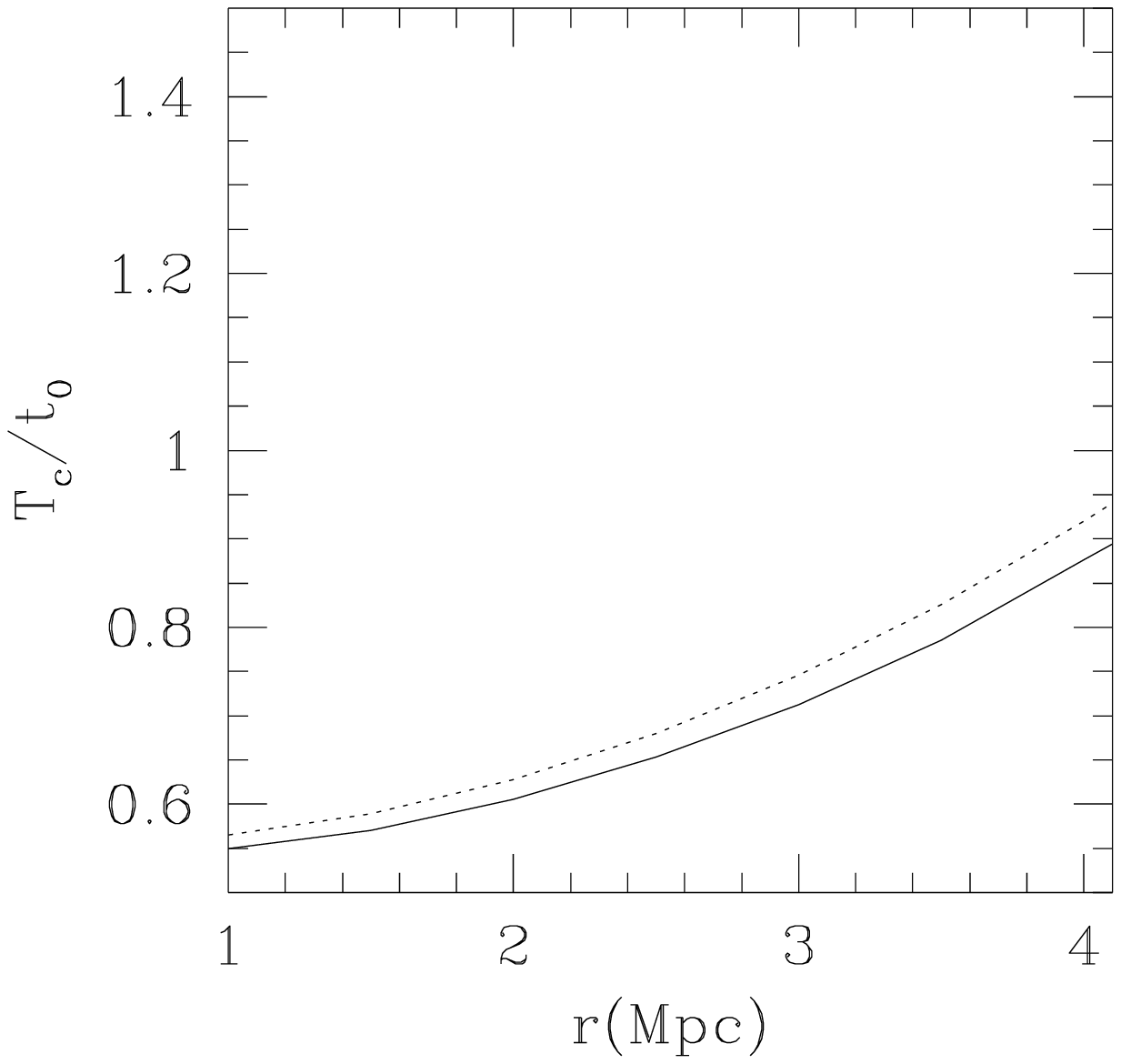,width=12cm}
\caption[]{The time of collapse of a shell of matter in units of the age of the
Universe $t_{0}$ for $\nu=3$ (dotted line) compared with GG's
model (solid line).}
\end{figure}
\begin{figure}[ht]
\psfig{file=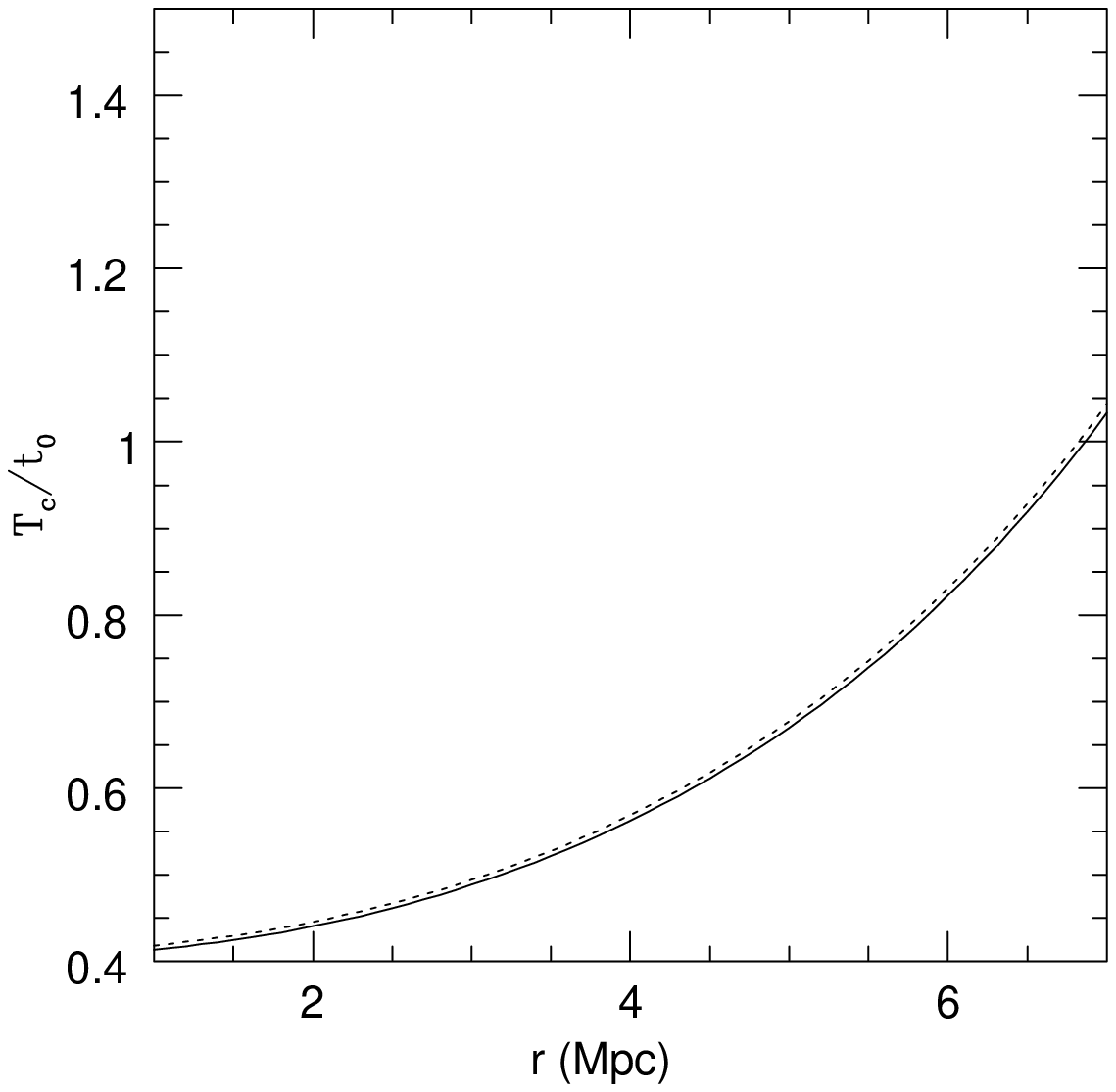,width=12cm}
\caption[]{The time of collapse of a shell of matter in units of the age of the
Universe $t_{0}$ for $\nu=4$ (dotted line) compared with GG's
model (solid line).}
\end{figure}
As shown the
presence of non-radial motions produces an increase in the time of collapse
of a spherical shell. The collapse delay is larger for a low value of $\nu $
and becomes negligible for $\nu \ge 3$. This result is in agreement with
the angular momentum-density anticorrelation effect: density peaks having
low value of $\nu $ acquire a larger angular momentum than high $\nu $ peaks
and consequently the collapse is more delayed with respect to high $\nu $
peaks. Given $ T_c(r,\nu )$ we also calculated the total mass gravitationally
bound to the final non-linear configuration. There are at least two criteria
to establish the bound region to a perturbation $\delta (r)$: a statistical
one (Ryden 1988b), and a dynamical one (Hoffman \& Shaham 1985). The
dynamical criterion, that we have used, assumes that the binding radius is
given by the condition that a mass shell collapse in a time, $ T_c$, smaller
than the age of the universe $ t_0$:
\begin{equation}
T_c(r,\nu )\leq t_0
\end{equation}
We calculated the time of collapse of GG spherical model, $ T_{c0}(r,\nu )$, 
using the density profiles given in Eq. (\ref{eq:dens}) for $ 1.7 < \nu < 4$ 
and then we repeated the calculation taking into account non-radial motions 
obtaining $T_c(r,\nu )$.
Then we calculated the binding radius, $r_{b}(\nu)$, for a GG
model solving $T_{c0}(r,\nu ) \leq t_0$ for $r$ and for several values
of $\nu$,
while we calculated the binding radius
of the model that takes into account non-radial motions,  $r_{b}(\nu)$,
repeating the calculation, this time with
$T_c(r,\nu )\leq t_0$.
We found a relation between $\nu $ and the mass of the cluster using the
equation: $M = \frac{4\pi }3r_b^3\rho _b$.\\
In Fig. 5 we compare the peak mass obtained from GG model, using
Hoffman \& Shaham's (1985) criterion, with that obtained from the model taking
into account non-radial motions.
\begin{figure}[ht]
\psfig{file=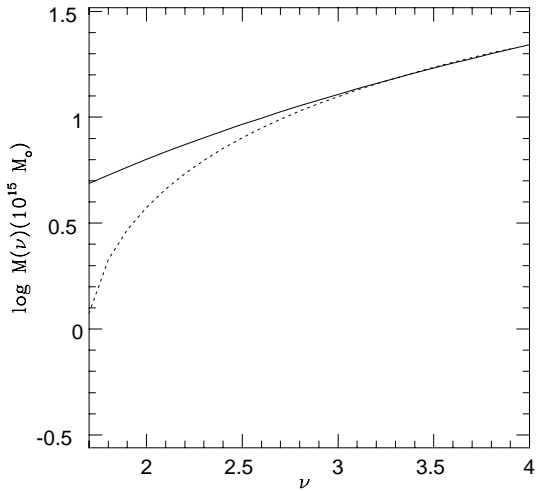,width=12cm}
\caption[]{The mass accreted by a collapsed perturbation, in units of
$10^{15}M_{\odot}$, taking into account non-radial motions 
effect (dotted line)  
compared to GG's mass (solid line).}
\end{figure}
As shown for high values of $\nu $ ($\nu \ge 3$), the two models give the
same result for the mass while for $\nu \leq 3$ the effect of non-radial motions
produces less bound mass with respect to GG model.\\
Before concluding this section we want to discuss the applicability of the spherical model and consequently the use of spherical shells in our model. 
The question of the applicablity of the idealized spherical collapse
model and that of the secondary infall model (SIM) to realistic systems and
initial conditions is almost as old as the model itself (GG;
Gunn 1977; Filmore \& Goldreich 1984; Quinn et al. 1986; Zurek et al. 1988; Quinn \& Zurek 1988; Warren et al. 1991;
Crone et al. 1994). In a recent paper Zaroubi et al. (1996) investigate the applicability of the quoted model to realistic
systems and initial conditions by comparing 
the results obtained using the
spherical model and SIM with that of
a set of simulations performed with
different N-body codes (Treecode and a "monopole term" code). They introduced
a numerical model to trace the evolution of density peaks 
under the assumption of aspherically symmetric force and realistic
initial conditions instead of the spherically symmetric force and initial
conditions as assumed in the SIM.
They obtained  a good agreement between the SIM 
and the simulations.

\section{Tidal field and the selection function}

According to biased galaxy formation theory the sites of formation of
structures of mass $\sim M$ must be identified with the maxima of the
density peak smoothed over a scale $R_f$ ($M\propto R_f^3$). A necessary
condition for a perturbation to form a structure is that it goes non-linearly
and that the linearly extrapolated density contrast reaches the value 
$\overline{\delta}(r)\geq \delta_c = 1.68$ 
or equivalently that the threshold criterion 
$\nu _t>\delta _c/\sigma_0(R_f)$ is satisfied, being $\sigma_0(R_f)$ 
the variance of the density field smoothed on scale $R_f$.
When these conditions are satisfied the matter in a shell around a peak
falls in toward the cluster center and virializes. In this scenario only
rare high $\nu $ peaks form bright objects while low $\nu $ peaks ($\nu
\approx 1$) form under-luminous objects. The kind of objects that forms from
non-linear structures depends on the details of the collapse. Moreover, if
structures form only at peaks in the mass distribution they will be more
strongly clustered than the mass.\\
Several feedback mechanisms have been
proposed to explain this segregation effect (Rees 1985; Dekel \& Rees 1987). 
Even if these feedback mechanisms work one cannot expect they have effect
instantaneously, so the threshold for structure formation cannot be sharp 
(BBKS). To
take into account this effect BBKS
introduced a threshold or selection
function, $t(\nu /\nu _t)$. The selection 
function, $t(\nu /\nu _t)$, gives the probability that a
density peak forms an object, while the threshold level, $\nu _t$,
is defined so that the probability that a peak forms an object is 1/2 when $
\nu =\nu _t$. 
The selection function introduced by BBKS (Eq. 4.13), is an empirical one
and depends on two parameters: the threshold $\nu _{t}$ and the shape
parameter $ q$:
\begin{equation}
t(\nu /\nu _t)=\frac{(\nu /\nu _t)^q}{1+(\nu /\nu _t)^q}
\end{equation}
If $q\rightarrow \infty $ this selection function is a Heaviside function $%
\vartheta (\nu -\nu _t)$ so that peaks with $\nu >\nu _t$ have a
probability equal to 100\% to form objects while peaks 
with $ \nu \leq \nu_{t}$ do not form objects. 
If $q$ has a finite value 
sub-$\nu _t$ peaks are selected with non-zero probability. Using the given
selection function the cumulative number density of peaks higher than $\nu $
is given, according to BBKS, by:
\begin{equation}
n_{pk}=\int_\nu ^\infty t(\nu /\nu _t)N_{pk}(\nu )d\nu 
\end{equation}
where $N_{pk}(\nu )$ is the comoving peak density 
(see BBKS Eq.~4.3).
A form of the selection function, physically founded, can be obtained
following the argument given in CAD. 
In this last paper the selection function is defined as:
\begin{equation}
t(\nu )=\int_{\delta _c}^\infty p\left[ \overline{\delta} ,
\langle \overline{\delta} (r_{Mt},\nu
 )\rangle ,\sigma _{\overline{\delta}} (r_{Mt},\nu )\right] d\delta \label{eq:sel}
\end{equation}
where the function 
\begin{equation}
p\left[ \overline{\delta} ,\langle \overline{\delta} (r) \rangle\right] = 
\frac 1{\sqrt{2\pi }\sigma_{\overline{\delta}} }\exp \left( -\frac{|\overline{\delta} -
\langle \overline{\delta} (r) \rangle|^2}{2\sigma
_{\overline{\delta}} ^2}\right) \label{eq:gau}
\end{equation}
gives the probability that the peak overdensity is
different from the average, in a Gaussian density field. The selection
function depends on $\nu $ through the dependence of $\overline{\delta} (r)$ 
on $\nu $. 
As displayed, the integrand is evaluated at a radius $r_{Mt}$ which is the
typical radius of the object that we are selecting. Moreover, the selection
function $t(\nu )$ depends on the critical overdensity threshold for the
collapse, $\delta _{c}$, which is not constant as in a spherical model
(due to the presence, in our analysis, of non-radial motions that delay the
collapse of the proto-cluster) but it depends on $\nu $. The dependence of 
$\delta _{c}$ on $\nu $ can be obtained in several ways, for example 
according to
Peebles (1980) the value of $ \delta _c$ depends on the ratio
$ T_c / t_m$ between the perturbation
collapse time, $T_c$, and 
its turn around time, $t_m$:
\begin{equation}
\delta _{c}=\frac 35\left( \frac{3\pi }4\frac{T_c}{t_m}\right)
^{2/3} 
\label{eq:del}
\end{equation}
Non-radial motions slow down the collapse of the mass shell with respect to
the GG collapse time changing the value of $T_{c}$. 
Using the calculated time of collapse for a given shell, and its dependence
on $\nu $, $\delta_{c}(\nu)$ can be calculated using Eq. (\ref{eq:del}).
An analityc determination of $ \delta_{c}(\nu)$ can be obtained following
a technique similar to that used by  
Bartlett \& Silk (1993). Using 
Eq.~(\ref{eq:ses}) it is possible to obtain the value of the
expansion parameter of the turn around epoch, $a_{max}$,
which is characterized by the condition $\frac{da}{dt}=0$. Using the relation
between $v$ and $\delta_{i}$,
in linear theory (Peebles 1980), we can find $ C$
that substituted in Eq. (\ref{eq:ses}) gives at turn around:
\begin{equation}
\delta_{c}(\nu)=\delta_{c0}
\left[1+\frac{8 G^{2}}{\Omega_{o}^{3} H_{o}^{6} r_{i}^{10}
\overline{\delta}(1+\overline{\delta})^{2}}
\int \frac{L^{2}da}{a^{3}}  \right]
\end{equation}
where $\delta_{c0}=1.68$ is the critical threshold for GG's model.
In Fig. 6 we show the
overdensity threshold as a function of $\nu $. As shown, $\delta _c(\nu )$
decreases with increasing $\nu $: when $ \nu > 3$ the threshold assumes the
typical value of the spherical model. This means, according
to the cooperative galaxy formation theory, (Bower et al. 1993)
that structures form more easily if there are other structures
nearby, i.e. the threshold level is a decreasing function of the mean mass
density.
\begin{figure}[ht]
\psfig{file=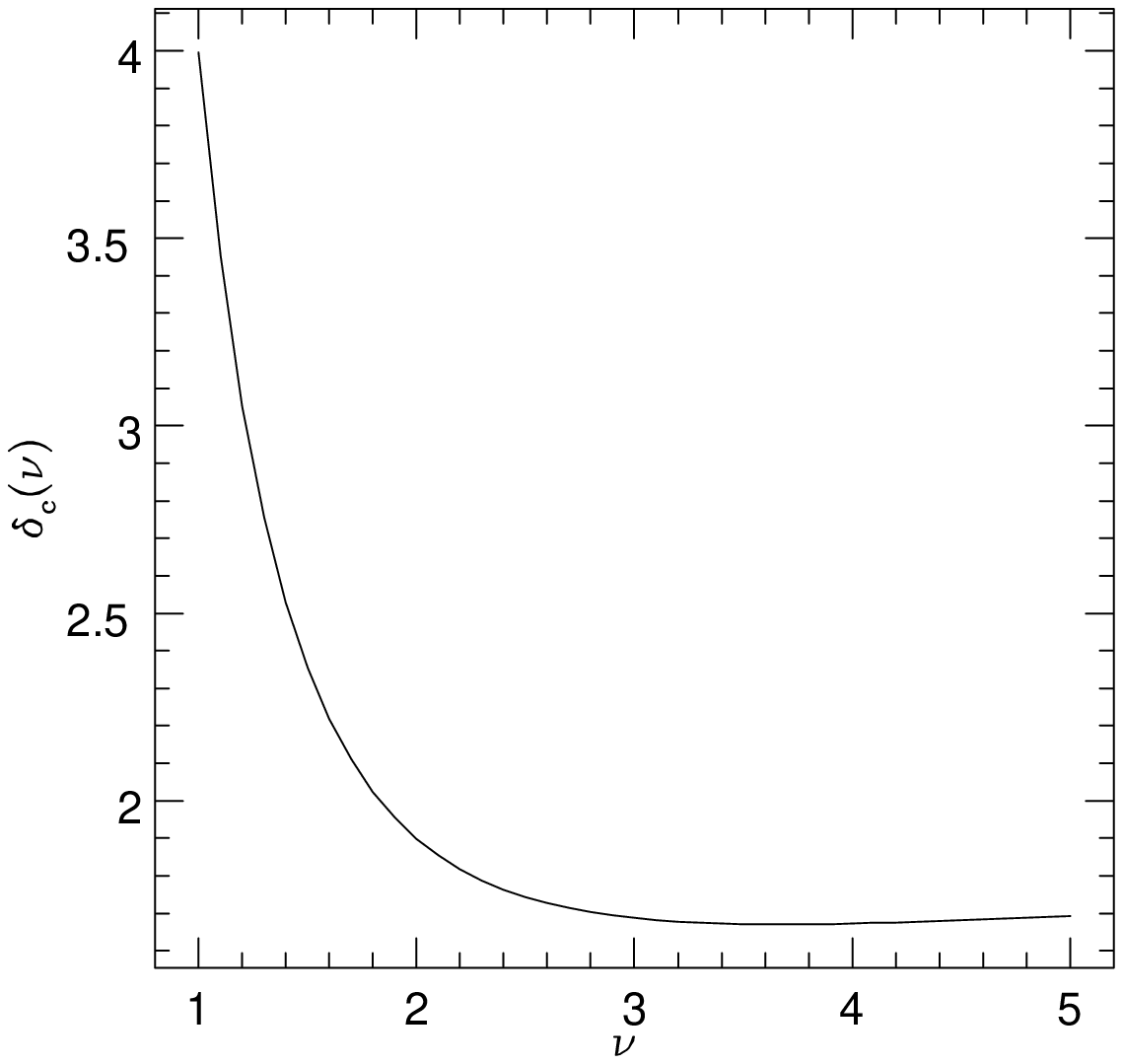,width=12cm}
\end{figure}
Known $\delta_{c}(\nu)$ and chosen a spectrum, the selection function is
immediately obtainable through Eq. (\ref{eq:sel}) and Eq. (\ref{eq:gau}).
The result of the calculation, plotted in Fig. 7, for two values
of the filtering radius, ($R_{f}=3$, $4$ $h^{-1}Mpc$),
shows that the selection
function, as expected, differs from an Heaviside function (sharp threshold).
\begin{figure}[ht]
\psfig{file=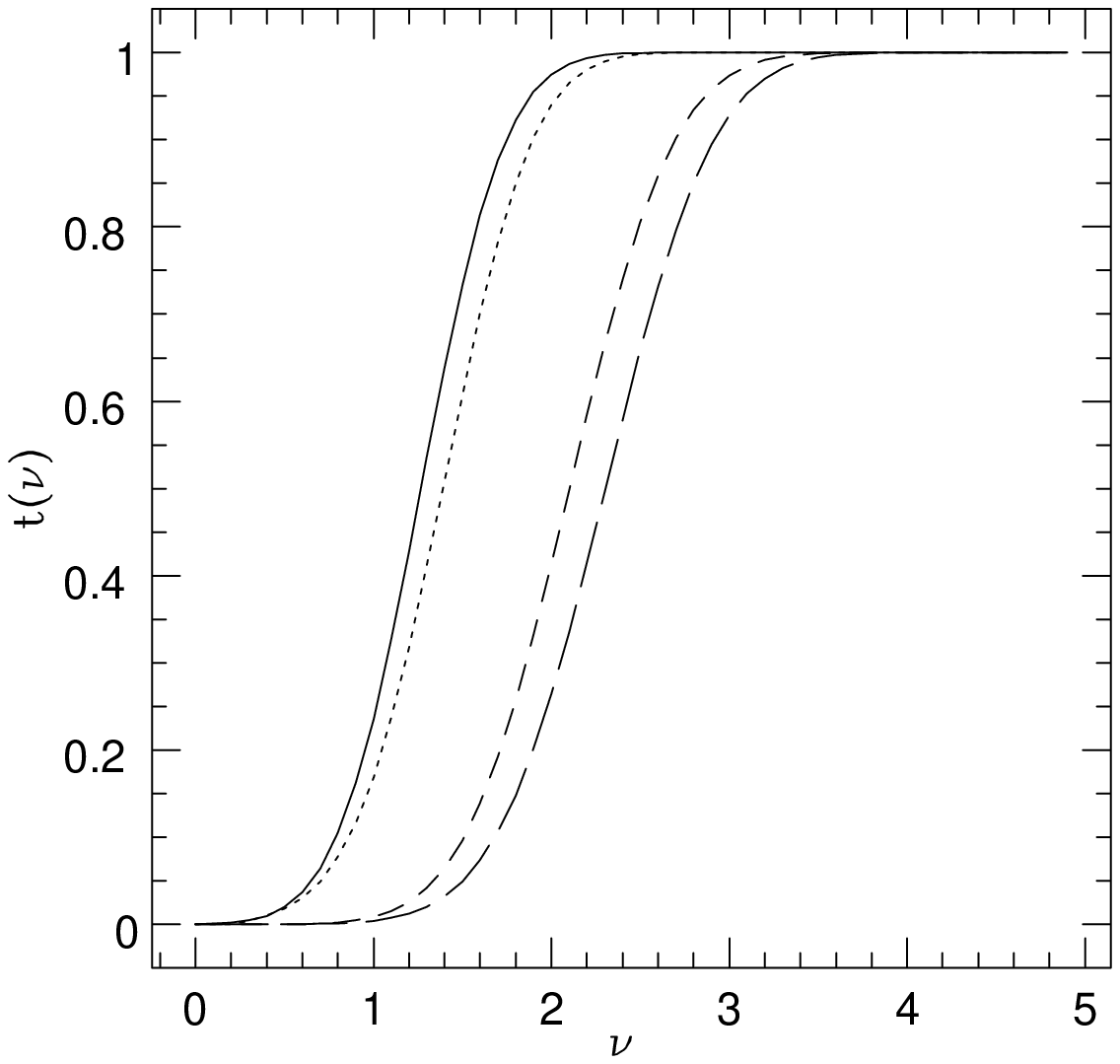,width=12cm}
\caption[]{The selection function, $t(\nu)$, for $R_{f}= 3h^{-1}Mpc$
(the solid line plots the selection function obtained without
taking into account the effects of non-radial motions;
the dotted line plots the selection function obtained taking into account
the effects of non-radial motions)
and for $4h^{-1}Mpc$
(the short dashed
line plots the selection function obtained without taking into account
the effects of non-radial motions;
the long dashed line plots the selection function obtained taking into account
the effects of non-radial motions).}
\end{figure}
The shape of the selection function depends on the values of the
filtering length $R_{f}$ and on non-radial motions.
The value of $\nu$ at which the selection function $ t(\nu)$ reaches the
value 1 ($t(\nu) \simeq 1$) increases for growing values of the filtering
radius, $R_{f}$. This is due to the smoothing effect of the filtering process.
The effect of non-radial motions is, firstly, that of shifting
$t(\nu)$ towards higher values of $\nu$, and, secondly, that of
making it steeper. 
The selection function is also different from that
used by BBKS (Tab. 3a). 
Finally it is interesting to note that the selection function defined by
Eq. (\ref{eq:sel}) and Eq. (\ref{eq:gau}) is totally general, it does not
depend on the presence or absence of non-radial motions. The latter
influences the selection function form through the change 
of $\delta_{c}$ induced by non-radial motions itself.

\section{The bias coefficient}

A model of the Universe in which light traces the mass distribution
accurately (unbiased model) is subject to several problems. As pointed
out by Davis et al. (1985) an unbiased CDM produces a galaxy correlation
function which is steeper than that observed and a pairwise velocity dispersion
larger than that deduced from redshift surveys. A remedy to this
problem can be found if we do not assume that light traces mass 
and adopt the biasing concept, i.e., galaxies are more clustered than
the distribution of matter in agreement with the concept of biasing inspired
by Kaiser (1984). The observations show that clusters of galaxies
cluster more strongly than galaxies, in the sense that the dimensionless
two-point correlation function, $\xi _{cc}(r)$, is much larger than the
galaxy two-point function, $ \xi _{gg}(r)$.
The galaxy two-point correlation function $ \xi _{gg}(r)$
is a power-law:
\begin{equation}
\xi_{gg}(r) = \left(\frac{r}{r_{0,g}}\right)^{\gamma}
\label{eq:unoventotto}
\end{equation}
with a correlation length $ r_{0,g} \simeq 5 h^{-1}$ Mpc and
a slope $ \gamma \simeq 1.8$ for
$ r \le 10 h^{-1}$ Mpc (Davis \& Peebles 1983; Davis et al. 1985; 
Shanks et al. 1989), (some authors disagree with this values; 
for example Strauss et al. 1992 and Fisher et al. 1993  
find $ r_{0,g} \simeq 3.79 h^{-1}$ Mpc and $ \gamma \simeq 1.57$). 
As regards the clusters of galaxies the form of the 
two-point correlation function, $\xi _{cc}(r)$, is equal to that given
by Eq. (\ref{eq:unoventotto}). Only the correlation length
is different. In the case
of clusters of galaxies the value of $ r_{0,c}$ is uncertain (see 
Bahcall \& Soneira 1983; Postman et al. 1986; Sutherland 1988; 
Bahcall 1988; Dekel et al. 1989; Olivier et al. 1990  and Sutherland \& 
Efstathiou 1991) however it lies in the range 
$  r_{0,c} \simeq 12 \div 25 h^{-1}$ Mpc in any case larger than $ r_{0,g}$.
One way of defining the bias coefficient of a class of objects
is that given by (BBKS):
\begin{equation}
b(R_f)= \frac {\langle \tilde\nu \rangle}{\sigma_0}+1
\end{equation}
where $ \langle \tilde\nu \rangle$ is:
\begin{equation}
\langle \tilde\nu \rangle = \int_0^\infty \left[ \nu -\frac{\gamma \theta }{1-\gamma ^2}\right] 
t\left(\frac{\nu}{\nu_{t}}\right) N_{pk}(\nu ) d\nu 
\label{eq:nu}
\end{equation}
We want to remember that, as shown by
Coles (1993), the biasing parameter can be also estimated by means of the
ratio of the amplitudes of the correlation function, $\xi(r)$, and the matter
auto-covariance function, $\Gamma(r)$:
\begin{equation}
b^ {2}(r)=\frac{\xi(r)}{\Gamma(r)}
\end{equation}
or by means of the ratio of the cumulative
integral of the two-points correlation function
($K_{3}(r)=\int_{0}^{r} \xi(r) r^{2}dr$) and that of the auto-covariance
function ($J_{3}(r)=\int_{0}^{r} \Gamma(r) r^{2}dr$):
\begin{equation}
b^{2}(r)=\frac{K_{3}(r) }{J_{3}(r)}
\end{equation}
 or finally the
ratio of galaxy, $Q(k)$, to mass, $P(k)$, power spectra:
\begin{equation}
b^ {2}(r)=\frac{Q(k)}{P(k)}
\end{equation}
 As stressed by Coles (1993) a local
bias generally produces a different response in each of these descriptors.
Even the qualitative behaviour of the limit of large scales can be different,
i.e. $ b(r)$ can increase or decrease. So one should decide very carefully 
which one of these definitions must be used, when discussing the
behaviour of the biasing parameter.
Being conscious of these difficulties we have chosen one of the most
popular descriptor of the biasing parameter (see also  Lilje 1990,
Liddle \& Lith 1993,
Croft \& Efstathiou 1994, CAD) in order to make comparisons with
other models.
From Eq. (\ref{eq:nu})
it is clear that the bias parameter can be calculated once a
spectrum, $P (k)$, is fixed. The bias parameter depends on the shape 
and normalization of the power spectrum. A larger value
is obtained
for spectra with more power on large scale (Kauffmann et al. 1996). 
In this calculation we continue to use the standard CDM
spectrum ($\Omega_0 = 1$, $ h = 0.5$) normalized imposing that the rms density
fluctuations in a sphere of radius $ 8 h^{-1}Mpc$ is the same as that observed
in galaxy counts, i.e. $\sigma _8=\sigma (8h^{-1}Mpc)=1$. The calculations
have been performed for three different values of the
filtering radius ($R_f=2,$ $ 3$, $ 4$
$h^{-1}Mpc$). The values of $ b$, that we have obtained, are respectively, 
in increasing order $ R_f$, 1.6, 1.93 and 2.25.\\
As shown,
the value of the bias parameter tends to increase with $R_f$ due the filter
effect of $ t(\nu)$. As shown $t(\nu)$ acts as a filter, increasing the
filtering radius, $R_{f}$, the value of $\nu$ at which
$t(\nu) \simeq 1$ increases. In other words when $ R_{f}$ increases, $t(\nu)$ selects
density peaks of a larger height. The reason for this behavior must be
searched in the smoothing effect that the increasing of the filtering radius
produces on density peaks. When $ R_{f}$ increases the density field
smooths and $ t(\nu)$ has to shift towards a higher value of $ \nu$
in order to select a class of object of fixed mass $ M$.

\section{Conclusions}

In this paper we have studied the role of non-radial motions on the collapse of
density peaks solving numerically the equations of motion of a shell of
barionic matter falling into the central regions of a cluster of galaxies.
We have shown that non-radial motions produce a delay in the collapse of
density peaks having a low value of $ \nu$ while the collapse of density peaks
having $ \nu > 3$ is not influenced. A first consequence of this effect
is a reduction of the mass bound to collapsed perturbations and an increase
in the
critical threshold, $ \delta_{c}$, which now is larger than that of the
top-hat spherical model and depends on $ \nu$. This means that
shells of matter of low density
have to be subject to a larger gravitational potential,
with respect to the homogeneous GG's model, in order to collapse. \\
The delay in the
proto-structures collapse gives rise to a dynamical bias similar to that
described in CAD whose bias
parameter may be obtained once a proper selection function is defined. 
The selection function found is not a pure Heaviside function and is
different from that used by BBKS to study the statistical
properties of clusters of galaxies. Its shape depends on the effect of
non-radial motions through its dependence on $\delta_{c}(\nu)$.
The function $t(\nu)$ selects higher and higher density peaks 
with increasing value of $ R_{f}$ due to the smoothing effect of the 
density field produced by the filtering procedure. 
Using this selection function and BBKS
prescriptions we have calculated the coefficient of bias $b$. \\
On clusters scales for $R_{f} = 4h^{-1}Mpc$ we found a value 
of $b= 2.25$
comparable with that obtained from the mean mass-to-light ratio of
clusters, APM survey, or from N-body simulations combined with
hydrodynamical models (Frenk et al. 1990).
Morever, the value of the coefficient of biasing $ b$ that we have calculated 
is comparable with the values of $ b$ given by Kauffmann et al. (1996). 
This means that non-radial
motions play a significant role in determining the bias level.
In our next paper (Del Popolo \& Gambera, in preparation) 
we make a more detailed analysis on the problem of the bias. 

\begin{flushleft}
{\it Acknowledgements}
\end{flushleft}
We would like to thank the anonymous referee for insightful comments
which led us to explain better our ideas. 
Besides, we are grateful to V. Antonuccio-Delogu  
for helpful and stimulating discussions during the period in which 
this work was performed.

\end{document}